\documentclass[english]{article}
\usepackage{amssymb}
\usepackage{fontenc}
\usepackage[latin1]{inputenc}
\usepackage{amsmath}
\usepackage{fontenc}
\usepackage{babel}

\input{tcilatex}

\begin{document}

\title{\textbf{Consistency restrictions on maximal electric field strength
in QFT}}
\author{S.P. Gavrilov\thanks{%
Department of General and Experimental Physics, Herzen State Pedagogical
University of Russia, Moyka emb. 48, 191186 St. Petersburg, Russia; e-mail:
gavrilovsergeyp@yahoo.com} and D.M. Gitman \thanks{%
Institute of Physics, University of São Paulo, CP 66318, CEP 05315-970 São
Paulo, SP, Brazil; e-mail: gitman@dfn.if.usp.br}}
\maketitle

\begin{abstract}
QFT with an external background can be considered as a consistent model only
if backreaction is relatively small with respect to the background. To find
the corresponding consistency restrictions on an external electric field and
its duration in QED and QCD, we analyze the mean-energy density of quantized
fields for an arbitrary constant electric field $E$, acting during a large
but finite time $T$. Using the corresponding asymptotics with respect to the
dimensionless parameter $eET^{2}$, one can see that the leading
contributions to the energy are due to the creation of paticles by the
electric field. Assuming that these contributions are small in comparison
with the energy density of the electric background, we establish the
above-mentioned restrictions, which determine,\emph{\ }in fact, the time
scales from above of depletion of an electric field due to the backreaction.

\end{abstract}

I. It is well-known that QFT in an external background provides an efficient
model for the study of quantum processes in those cases when some part of a
quantized field is strong enough to be treated as a classical one. For
example, QED with an external electromagnetic background formally arises
from extending the QED Lagrangian by the interaction of the matter current
with a given external electromagnetic field $A_{\mu }^{\mathrm{ext}}$, which
is not quantized. This is naturally implied as a certain approximation. The
study of some problems in QED and QCD with superstrong external backgrounds
and their applications to astrophysics and condensed matter (to graphene
physics) has once again raised the question of a consistency of such
theories. Obviously, the question must be answered, first of all, in the
case of a constant external field. Calculations that have an immediate
relation to the above-mentioned problem have first been carried out by
Heisenberg and Euler in the case of QED with constant parallel electric $E$
and magnetic $B$ fields, see \cite{HeiE36}. They computed the change of the
vacuum energy of spinning particles for arbitrary $B$ and a weak electric
field $E\ll E_{c}=m^{2}/e$ ($\hslash =c=1$) which is unable to effectively
create pairs from vacuum. They interpreted this change as a change of the
energy of the external field itself and, at the same time, as a change of
the Maxwell Lagrangian $\mathcal{L}^{(0)}=\left( E^{2}-B^{2}\right) /8\pi $
by a certain addition $\mathcal{L}^{(1)}$. The limiting case of a strong
magnetic field ($B\gg m^{2}/e,\;E=0$) yields 
\begin{equation}
\mathcal{L}^{(1)}\approx -\left( \frac{\alpha }{3\pi }\ln \frac{eB}{m^{2}}%
\right) \mathcal{L}^{(0)}\,,  \label{3}
\end{equation}%
where $\alpha =e^{2}\ $is\ the fine-structure\ constant. This result is in
agreement with more advanced calculations carried out by Ritus \cite{Rit75},
who arrived at the conclusion that the loop expansion makes sense only for
the magnetic fields restricted by the condition $B\ll F_{\max }~$,%
\begin{equation*}
F_{\max }=\frac{m^{2}}{e}\exp \left( \frac{3\pi }{\alpha }\right) \approx 
\frac{m^{2}}{e}10^{560}\,.
\end{equation*}

Shabad and Usov \cite{ShaU05} have recently established a more rigid
limitation for the maximal admissible strength of magnetic field, $B\ll
10^{28}m^{2}/e$,\textrm{\ }having analyzed the structure of QED of vacuum in
this field, taking into account the interaction of virtual electron-positron
pairs.

The addition $\mathcal{L}^{(1)}$\ has been generalized in a certain way\emph{%
\ }to an arbitrary constant field (to arbitrary $E$) and is now called the
Heisenberg-Euler Lagrangian (HEL), for a review, see \cite{Dunn04}. However,
its physical meaning for a strong electric field is not completely clear.
The problem has been approached from another angle by Schwinger \cite{S51},
who supposed that the variation of an effective Lagrangian of
electromagnetic field should arise due to a non-zero (in external fields) $%
in-out$ vacuum current of charged particles, without any restrictions on the
intensity of electric field; in doing so, he presented its calculation in a
constant external field, and obtained precisely the HEL. In the general
case, the HEL is complex-valued, its imaginary part determines the
probability of pair-creation, as has been confirmed by independent
calculations \cite{Nikis79,FraGi81}. Considering the $in-out$ vacuum
current, Schwinger has made it possible to obtain an elegant expression for
his effective Lagrangian in terms of the causal (Feynman's) Green function.
Nevertheless, such an effective Lagrangian is not related to the problem of
mean values, in particular, it does not reproduce the change of the vacuum
energy of spinning particles for arbitrary $E$; the latter problem has to be
formulated independently as a mean-value problem, and is expressed via
noncausal Green's functions; see \cite{FraGi81}.

In a strong electric field ($E\gg m^{2}/e,\;B=0$), the real-valued part of
HEL, describing the effects of vacuum polarization, is given by the{\LARGE \ 
}right-hand side of (\ref{3}), where $B$ is replaced by $E$. From this
expression one can extract the negative-valued additive correction $\mathcal{%
E}^{\left( 1\right) }=\func{Re}\,\mathcal{L}^{(1)}$ to the classical Maxwell
density of energy, $\mathcal{E}^{\left( 0\right) }=\mathcal{L}^{(0)}$.
Consequently, in order that the total energy density of electric field $%
\mathcal{E}=\mathcal{E}^{\left( 0\right) }+\mathcal{E}^{\left( 1\right) }$
should be zero, the only effects of vacuum polarization at the value $E\sim
F_{\max }$ are sufficient by themselves. On these grounds, in \cite{GreRo73}
it was suggested that $F_{\max }$ should be also a limiting value of
electric field. It turns out that $\func{Im}L^{(1)}\sim E^{2},$ which\ can
be interpreted as an evidences that the vacuum instability is less than the
vacuum polarization. This is not true since the vacuum instability is a
nonlocal effect, being directly dependent on the electric field duration.

In our opinion, the most adequate object, whose analysis can answer these
questions, is the mean value of the energy-momentum tensor of matter,
computed with respect to various initial states. A detailed calculation of
such a mean value in QED in the one-loop approximation, taking an exact
account of the interaction with the electric background, has been given in 
\cite{GavG07}. Here, we will only use some of these results, in particular,
the mean energy density of matter for large values of strength and duration
of the electric field. We examine two cases, when the initial state of the
quantized Dirac field is vacuum, and is in thermal equilibrium.\ In
addition, we consider the case of charged bosons\ and QCD with an external
chromoelectric field. We demonstrate that under these conditions, the effect
of particle-creation is precisely the main reason for the change of the
energy of matter. Making a comparison between the change of the energy
density of matter and the energy density of the external electric field,
which is responsible for this change, we obtain restrictions on the
intensity of the external field and its duration, which we call consistency
restrictions.

II. A constant electric field acting during an infinite time creates an
infinite number of pairs from vacuum even in a finite volume. This is why we
choose{\Large \ }the external background as a constant electric field acting
during a finite period of time; we refer to this field as a $T$-constant
field. The finiteness of field duration is a natural regularization in the
given problem; on the other hand, it is a necessary physical parameter,
which subsequently enters the consistency restriction on the value of the
maximal electric field. The\ $T$-constant field turns on at $%
-T/2=t_{1}>t_{in}$ and turns off at $T/2=t_{2}<t_{out}$. We choose the
nonzero $T$-constant field potential $A_{3}(t)$ as a continuous function of
the form $A_{3}(t)=-Et$ for $t\in \lbrack t_{1},t_{2}]$, being constant for $%
t\in (-\infty ,t_{1})$ and $t\in (t_{2},+\infty ).$ The effects of
particle-creation by the $T$-constant field have been studied in detail in 
\cite{GavG96a}. In particular, it was shown that in case%
\begin{equation}
T\gg \left( eE\right) ^{-1/2}\left[ 1+m^{2}/eE\right]  \label{10}
\end{equation}%
all the finite effects caused by particle-creation reach their asymptotic
values, whereas the details involving the form of switching the field on and
off can be neglected. In our calculations, we assume this restriction from
below for the time $T.$

Taking the time instant $t_{2}-0=T/2-0$ immediately before the electric
field has been turned off, we shall now examine the mean energy value $%
\langle \hat{H}\rangle $ of the spinor field on condition that its state at
the initial instant $t_{in}\rightarrow -\infty $ should be vacuum. The
potentials of the $T$-constant field do not depend on the spatial
coordinates, which implies that $\langle \hat{H}\rangle $ is proportional to
the space volume $V$. In the one-loop approximation, $\langle \hat{H}\rangle
=wV$, with the mean energy density $w$ being independent of the spatial
coordinates,%
\begin{equation}
w=\left. \frac{1}{2}\langle 0,in|\left[ \psi (x)^{\dagger },\mathcal{H}\psi
(x)\right] |0,in\rangle \right| _{x^{0}=t_{2}-0}\mathbf{~,}  \label{11}
\end{equation}%
where $\mathcal{H}$ is the one-particle Dirac Hamiltonian; $\psi (x)$ are
the operators of the Dirac field in the generalized Furry representation
(see, e.g., \cite{FraGi81}) obeying the Dirac equation with the external
background; $|0,in\rangle $ is the initial vacuum state in the same
representation. The above-mentioned choice allows one to take a complete
account of the pair-creation effect during the entire time. Also, since the
electric field has not yet been switched off, this allows us to make a
complete study of\emph{\ }the vacuum polarization effect. Notice that the
initial vacuum $\left| 0,in\right\rangle $ is identical with the vacuum of
those free particles that correspond to the initial potential $A_{3}=ET/2$.

The expression for $w$ is obviously real-valued. One can see that it can be
represented as%
\begin{equation}
w=-\frac{1}{4}\left. \left[ \lim_{t\rightarrow t^{\prime }-0}\mathrm{tr}%
\left[ \left( \partial _{0}-\partial _{0}^{\prime }\right)
S_{in}(x,x^{\prime })\right] +\lim_{t\rightarrow t^{\prime }+0}\mathrm{tr}%
\left[ \left( \partial _{0}-\partial _{0}^{\prime }\right)
S_{in}(x,x^{\prime })\right] \right] \right\vert _{\mathbf{x=x}^{\prime
},x^{0}=t_{2}-0}\,,  \label{12}
\end{equation}%
where $\mathrm{tr}\left[ \cdots \right] $ is the trace in the space of $%
4\times 4$ matrices, and $S_{in}(x,x^{\prime })$ is the so-called $in-in$
Green function,%
\begin{equation}
S_{in}(x,x^{\prime })=i\langle 0,in|T\psi (x)\bar{\psi}(x^{\prime
})|0,in\rangle =S^{c}(x,x^{\prime })+S^{p}(x,x^{\prime })\,,  \label{14}
\end{equation}%
where $S^{c}(x,x^{\prime })$ is Feynman's causal Green function, while the
function $S^{p}(x,x^{\prime })$ is a difference of two Green's functions,
satisfying the homogeneous Dirac equation; see \cite{FraGi81}. The final
vacuum $\left\vert 0,out\right\rangle $ is the vacuum of free particles in
the generalized Furry picture and corresponds to the constant potential $%
A_{3}=-ET/2$.

The separation of $S_{in}$ into the $c$- and $p$-parts is responsible for
the separation of $w$ into the two respective summands $w=w^{c}+w^{p}$. One
can verify that the expression for $w^{c}$ has a finite limit at $%
T\rightarrow \infty $, i.e., it permits a transition to the limit of a
constant electric field. Then $S^{c}$\ can be presented by a proper-time
integral, see \cite{Nik70}.  Using this expression, one can readily verify
that $w^{c}$ is expressed in terms of the real-valued part of HEL (at $B=0$%
). This contribution is due to vacuum polarization. In a superstrong
electric field, it has the form%
\begin{equation*}
w^{c}=E\frac{\partial \func{Re}\mathcal{L}^{(1)}}{\partial E}-\func{Re}%
\mathcal{L}^{(1)}\approx -\left( \frac{\alpha }{3\pi }\ln \frac{eE}{m^{2}}%
\right) \mathcal{L}^{(0)}.
\end{equation*}

The contribution $w^{p}$ arises due to particle-creation. It is computed as
follows. First of all, using the general theory of particle-creation (see, 
\cite{FraGi81}), one can represent the function $S^{p}(x,x^{\prime })$ in
the form%
\begin{equation}
\ S^{p}(x,x^{\prime })=i\sum_{nm}\,_{-}{\psi }_{n}(x)\,\left[
G(_{+}|^{-})G(_{-}|^{-})^{-1}\right] _{nm}^{\dagger }{_{+}\bar{\psi}}%
_{m}(x^{\prime })\,.  \label{17}
\end{equation}%
Here, $\left\{ _{\pm }\psi _{n}(x)\right\} $ are the so-called $in$%
-solutions of the Dirac equation in a $T$-constant electric field, their
asymptotics at $t\leq t_{1}$ being stationary states of free electrons $%
\left( +\right) $ and positrons ($-$) for the Dirac Hamiltonian with the
constant potential $A_{3}=ET/2$. The matrices $G\left( _{\pm }|^{\pm
}\right) $ (being a matrix generalization of the Bogolyubov coefficients)
are defined by decompositions of the so-called $out$-solutions in the $in$%
-solutions: $^{\pm }\psi (x)=_{+}\psi (x)G\left( _{+}|^{\pm }\right)
+_{-}\psi (x)G\left( _{-}|^{\pm }\right) .$ Here, $\left\{ ^{\pm }\psi
_{n}(x)\right\} $ are the $out$-solutions of the Dirac equation in a $T$%
-constant electric field; their asymptotics at $t\geq t_{2}$ describe free
particles (electrons $\left( +\right) $ and positrons ($-$)) with an energy
spectrum defined by the Dirac Hamiltonian with the constant potential $%
A_{3}=-ET/2$. The matrices $G\left( _{\pm }|^{\pm }\right) $ are expressed
via the inner products of $in$- and $out$-solutions, and obey some unitary
relations following from normalization conditions for these solutions.

In a $T$-constant uniform electric field, we can choose the quantum numbers
of particles as $n=\left( \mathbf{p},r\right) $, where $\mathbf{p}$ is the
particle momentum and $r=\pm 1$ is the spin projection; then the matrices $%
G\left( _{\pm }|^{\pm }\right) $ are diagonal, $G\left( _{\pm }|^{\pm
}\right) _{n,n^{\prime }}=\delta _{r,r^{\prime }}\delta _{\mathbf{p},\mathbf{%
p}^{\prime }}g\left( _{\pm }|^{\pm }\right) _{\mathbf{p},r}\ .$ Here,{\LARGE %
\ }we use{\LARGE \ }the standard volume regularization, so that $\delta (%
\mathbf{p}-\mathbf{p}^{\prime })\rightarrow \delta _{\mathbf{p},\mathbf{p}%
^{\prime }}$. In addition, the differential mean numbers of electrons (equal
to the corresponding differential mean number of pairs) created from vacuum
with a given momentum $\mathbf{p}$ and spin projections $r$ are given by $%
\aleph _{\mathbf{p},r}=\left\vert g\left( _{-}|^{+}\right) \right\vert ^{2}.$
It is easy to verify, using (\ref{17}), that the function $S^{p}$, which
enters the expression for $w^{p}$ at $x\approx x^{\prime }$, can be taken
there in the form%
\begin{equation*}
S^{p}(x,x^{\prime })=-i\int d\mathbf{p}\sum_{r=\pm 1}\aleph _{\mathbf{p},r}%
\left[ ^{+}{\psi }_{\mathbf{p},r}(x)\ ^{+}{\bar{\psi}}_{\mathbf{p}%
,r}(x^{\prime })-\,^{-}{\psi }_{\mathbf{p},r}(x)\ ^{-}{\bar{\psi}}_{\mathbf{p%
},r}(x^{\prime })\right] \,,~~x\approx x^{\prime }\ .
\end{equation*}%
For $t\geq t_{2}$, the $out$-solutions $^{\pm }\psi _{\mathbf{p},r}\left(
x\right) $ describe free particles with the quantum numbers $\mathbf{p},r$
and energies $\varepsilon _{\mathbf{p},r}=\sqrt{m^{2}+\mathbf{p}_{\bot
}^{2}+\left( eET/2-p_{3}\right) ^{2}}\,,\ \ \mathbf{p}_{\bot
}=(p^{1},p^{2},0).$ Taking all the above into account, we obtain%
\begin{equation}
w^{p}=\frac{1}{4\pi ^{3}}\int d\mathbf{p}\sum_{r=\pm 1}\aleph _{\mathbf{p}%
,r}\varepsilon _{\mathbf{p},r}\ .  \label{22}
\end{equation}%
This quantity is the mean energy density of pairs created from vacuum. It
can be estimated in the case of strong electric fields, $E\gtrsim E_{c}$ and
a sufficiently large $T$ as follows. As has been demonstrated in \cite%
{GavG96a}, in case the time $T$ is sufficiently large,%
\begin{equation*}
T\gg \left( m^{2}+\mathbf{p}_{\bot }^{2}+eE\right) \left( eE\right) ^{-3/2}\
,
\end{equation*}%
and the longitudinal momenta are restricted by the condition $|p_{3}|\leq
\left( \sqrt{eE}T/2-K_{p}\right) \sqrt{eE}$, where $K_{p}$ is a sufficiently
large arbitrary constant, $\sqrt{eE}T\gg K_{p}\gg \ 1+\left( m^{2}+\mathbf{p}%
_{\bot }^{2}\right) /eE$, the differential mean numbers $\aleph _{\mathbf{p}%
,r}$ have the form%
\begin{equation*}
\aleph _{\mathbf{p},r}^{\mathrm{asy}}=\exp \left( -\pi \frac{m^{2}+\mathbf{p}%
_{\bot }^{2}}{eE}\right) \,.
\end{equation*}%
For any fixed $\mathbf{p}_{\bot }^{2}$, the function $\aleph _{\mathbf{p},r}$
is fast-decreasing for $|p_{3}|>\left( \sqrt{eE}T/2-K_{p}\right) \sqrt{eE}$.
For this reason, we can disregard the contribution to the integral (\ref{22}%
) due to the integration over such momenta $p_{3}$ in comparison with the
main contribution, which is defined by the dimensionless parameter $eET^{2}$%
{\LARGE .} The latter parameter, in fact, determines a large integration
domain over $p_{3}$. In its turn, the exponential decrease of $\aleph _{%
\mathbf{p},r}^{\mathrm{asy}}$ with the grows of $\mathbf{p}_{\bot }^{2}$
allows one to ignore the contributions to the integral (\ref{22}) due to%
\emph{\ }a large $\mathbf{p}_{\bot }^{2}/eE$ $\gtrsim \sqrt{eE}T$.
Consequently, in order to evaluate the term which leads in $\sqrt{eE}T$ in
integral (\ref{22}) we can replace $\aleph _{\mathbf{p},r}$ by $\aleph _{%
\mathbf{p},r}^{\mathrm{asy}}$ under condition\emph{\ }(\ref{10}), while
restricting the limits of integration over momenta by the region $%
|p_{3}|\leq \sqrt{eE}\left( \sqrt{eE}T/2-K\right) $, where $K$ is a
sufficiently large arbitrary constant, $\sqrt{eE}T\gg K\gg 1+m^{2}/eE$%
\thinspace . Having calculated the integral (\ref{22}) over $\mathbf{p}%
_{\bot }$, we obtain the $T$-leading term in the form%
\begin{equation}
w^{p}=eET\aleph \ ,\ ~\aleph =\frac{e^{2}E^{2}T}{4\pi ^{3}}\exp \left( -\pi 
\frac{m^{2}}{eE}\right) \ .  \label{27}
\end{equation}

III. We now suppose that the energy density of particles $w=w^{p}$, arising
precisely due to the action of a $T$-constant electric field, should be
essentially smaller than the density of the electric field itself, being
equal to the classical Maxwell density of energy, $\mathcal{E}^{\left(
0\right) }=E^{2}/8\pi $. Thus, the condition of a smallness of back-reaction
is $w^{p}\ll E^{2}/8\pi ,$ which, owing to (\ref{27}), takes the form of a
restriction from above on the dimensionless parameter $eET^{2}$:%
\begin{equation}
eET^{2}\ll \frac{\pi ^{2}}{2e^{2}}\,\exp \left( \pi \frac{m^{2}}{eE}\right) .
\label{31}
\end{equation}%
On the other hand, all the asymptotic formulas have been obtained under
condition (\ref{10}), which restricts the mentioned parameter from below, $%
\left[ 1+m^{2}/eE\right] ^{2}\ll eET^{2}.$ Since $\pi ^{2}/2e^{2}\gg 1$,
there exists a region of values of $E$ and $T$ that satisfies both the
inequalities. We note that time scale from above in (\ref{31}) is more
restrictive than the scale derived from the rate of pair production, see 
\cite{Mad08}.

In case the initial state is in thermal equilibrium at temperature $\theta $%
, the mean energy density has an additional term $w_{\theta }^{c}$, which
represents, in fact, the work of a $T$-constant field on particles from the
initial state, as well as the term $w_{\theta }^{p}$%
\begin{equation*}
w_{\theta }^{p}=-\frac{1}{4\pi ^{3}}\int d\mathbf{p}\sum_{r=\pm 1}\aleph _{%
\mathbf{p},r}n_{\mathbf{p},r}\left( in\right) \varepsilon _{\mathbf{p},r}~,\
\ n_{\mathbf{p},r}\left( in\right) =\left[ \exp \left( \tilde{\varepsilon}_{%
\mathbf{p},r}/\theta \right) +1\right] ^{-1}\,,
\end{equation*}%
where $\tilde{\varepsilon}_{\mathbf{p},r}=\sqrt{m^{2}+\mathbf{p}_{\bot
}^{2}+\left( qET/2+p_{3}\right) ^{2}}$ is the energy of a free $in$\emph{-}%
particle.\emph{\ }The latter term determines a temperature-dependent
correction to the energy of particles created from vacuum; see \cite{GavG07}%
. The energies of particles that contribute to $w_{\theta }^{c}$ in the
limit of a large $T$ are mostly determined by a large longitudinal kinetic
momentum, with the energy being of order $eET$, as well as the energies of
particles created at $\theta =0$ in the expression (\ref{27}) for $w^{p}$.
Given that, however, the density of initial particles is constant, being
determined only by the initial condition, whereas the density of created
particles increases in proportion with $T$. Therefore, at large $T$ and $E,$ 
$w_{\theta }^{c}$ can be neglected in comparison with $w^{p}$, and $w\simeq
w^{p}+w_{\theta }^{p}\ .$

In case the initial state is in thermal equilibrium at low temperatures\emph{%
\ }$\theta \ll eET$,\emph{\ }the contribution\emph{\ }$w_{\theta }^{p}$\emph{%
\ }turns out to be small in comparison with\emph{\ }$w^{p}$\emph{. }At high
temperatures\emph{\ }$\theta \gg eET,$ the energy density has the form\emph{%
\ }$w=\left( eET/6\theta \right) w^{p}$\emph{\ .} Thus, the restriction (\ref%
{31}) is valid both for the vacuum initial state and for a low-temperature
initial thermal state. At high temperatures we have a weaker\emph{\ }%
restriction:%
\begin{equation}
\frac{\left( eE\right) ^{2}T^{3}}{\theta }\ll \frac{3\pi ^{2}}{e^{2}}\exp
\left( \pi \frac{m^{2}}{eE}\right) .  \label{33}
\end{equation}

Analogously, one can find restrictions for QED with charged bosons in a $T$%
-constant electric field. At low temperatures, we have%
\begin{equation*}
eET^{2}\ll \frac{\pi ^{2}}{Je^{2}}\,\exp \left( \pi \frac{m^{2}}{eE}\right) ,
\end{equation*}%
where $J$ is the number of the spin degrees of freedom ($J=1$ for scalar
particles and $J=3$ for vector particles). In the case of high temperatures
the restriction has a completely different character than (\ref{33}), namely,%
\begin{equation*}
\theta T\ln \left( \sqrt{eE}T\right) \ll \frac{\pi ^{2}}{2Je^{2}}\,\exp
\left( \pi \frac{m^{2}}{eE}\right) .
\end{equation*}%
One can easily extend these results to $D+1$\ dimensions, using the
corresponding $N$ in (\ref{27}), taken from Eq. (37) in \cite{GavG96a}.

IV. A similar analysis can be performed in the case of QCD with an
electric-like non-Abelian external background. Such a background is a part
of the known chromoelectric flux-tube model \cite{CasNN79}. At present, the
chromoelectric field is associated \cite{KhaLT06} with an effective theory,
color glass condensate. Here, we shall derive restrictions on the external
background which allows one to treat particles created from vacuum still as
weakly coupled,{\large \ }owing to the property of asymptotic freedom in
QCD. To this end, we use the results obtained in \cite{GavGT06,NayN05} for
QCD with a constant $SU(3)$ chromoelectric field $E^{a}$ ($a=1,\ldots ,8$).
If the initial state is vacuum, the density of created gluons is noticeably
higher than the density of created quarks at any intensity of a $T$-constant
chromoelectric field; see \cite{GavGT06}.{\large \ }The same is valid at any
finite temperature; therefore, for our purposes it is sufficient to take
into account only the gluon contribution. It has been demonstrated in \cite%
{GavGT06} that the $\mathbf{p}_{\perp }$-distribution density $n_{\mathbf{p}%
_{\perp }}^{gluon}$ of gluons produced from vacuum with all the possible
values $p_{3}$ and the quantum numbers that characterize the inner degrees
of freedom can be presented as follows:%
\begin{equation}
n_{\mathbf{p}_{\perp }}^{gluon}=\frac{1}{4\pi ^{3}}\sum_{j=1}^{3}Tq\tilde{E}%
_{(j)}\aleph _{\mathbf{p}}^{(j)},\;\;\aleph _{\mathbf{p}}^{(j)}=\exp \left( -%
\frac{\pi \mathbf{p}_{\bot }^{2}}{q\tilde{E}_{(j)}}\right) ,  \label{39}
\end{equation}%
where $\tilde{E}_{(j)}$ are positive eigenvalues of the matrix $%
if^{abc}E^{c} $ for the adjoint representation of $SU(3)$, and $q$ is the
coupling constant. The $(j)$-terms in (\ref{39}) can be interpreted as those
obtained for Abelian-like electric fields $\tilde{E}_{(j)}$, respectively.
Then, the total energy density of gluons created from vacuum by the field $%
\tilde{E}_{(j)}$ is determined by integrals of the kind (\ref{22}). Taking
into account that maxima of the fields are restricted by the condition $%
\tilde{E}_{(j)}\leq \sqrt{C_{1}}$ ($C_{1}=E^{a}E^{a}$ is a Casimir invariant
for{\large \ }$SU(3)$) and the relation $\sum_{j=1}^{3}\tilde{E}%
_{(j)}^{2}=3C_{1}/2$, one can find that at low temperatures $\theta \ll q%
\sqrt{C_{1}}T$ the consistency restriction for the dimensionless parameter $q%
\sqrt{C_{1}}T^{2}$ has the form%
\begin{equation*}
q\sqrt{C_{1}}T^{2}\ll \pi ^{2}/3q^{2}\,.
\end{equation*}%
As in the case of QED, this restriction must be accompanied by a restriction
from below, $1\ll q\sqrt{C_{1}}T^{2}$, which is related to the fact that all
the asymptotic expressions have been obtained for sufficiently large values
of $T$. Therefore, the $T$-constant $SU(3)$ chromoelectric field
approximation is consistent during the period when the produced partons can
be treated as weakly coupled, due to the property of asymptotic freedom in
QCD.\textrm{\ }At high temperatures, $\theta \gg q\sqrt{C_{1}}T$, the
consistency restriction is far more rigid:%
\begin{equation*}
\theta T\ln \left( q\sqrt{C_{1}}T^{2}\right) \ll \pi ^{2}/3q^{2}\,.
\end{equation*}

The above established consistency restrictions determine, in fact, the time
scales from above of depletion of an electric field due to the backreaction.

\subparagraph{\protect\large Acknowledgement}

S.P.G. thanks FAPESP for support and Universidade de São Paulo for
hospitality. D.M.G. acknowledges the permanent support of FAPESP and CNPq.


\begin{thebibliography}{99}
\bibitem{HeiE36} W. Heisenberg and H. Euler, Z. Phys. \textbf{98}, 714
(1936).

\bibitem{Rit75} V. I. Ritus, Sov. Phys. JETP \textbf{42,} 774 (1975); ibid 
\textbf{46, }423 (1977).

\bibitem{ShaU05} A.E. Shabad and V.V. Usov, Phys. Rev. Lett. \textbf{96,}
180401 (2006).

\bibitem{Dunn04} G.V. Dunne, \emph{Heisenberg-Euler effective Lagrangians:
Basics and extensions}, in I. Kogan Memorial Volume, \emph{From fields to
strings: Circumnavigating theoretical physics}, Eds. M Shifman, A.
Vainshtein and J. Wheater, World Scientific, 2005; arXiv:hep-th/0406216.

\bibitem{S51} J. Schwinger, Phys. Rev. \textbf{82}, 664 (1951).

\bibitem{Nikis79} A.I. Nikishov, in \emph{Quantum Electrodynamics of
Phenomena in Intense Fields}, Proc. P.N. Lebedev Phys. Inst. \textbf{111},
153 (Nauka, Moscow 1979).

\bibitem{FraGi81} E.S. Fradkin and D.M. Gitman, Fortschr. Phys. \textbf{29,}
381 (1981); E.S. Fradkin, D.M. Gitman and S.M. Shvartsman, \emph{Quantum
Electrodynamics with Unstable Vacuum} (Springer-Verlag, Berlin 1991).

\bibitem{GreRo73} M. Greenman and F. Rohrlich, Phys. Rev. D \textbf{8}, 1103
(1973).

\bibitem{GavG07} S.P. Gavrilov and D.M. Gitman, arXiv:0709.1828 (accepted
for publication in Phys. Rev. D).

\bibitem{GavG96a} S.P. Gavrilov and D.M. Gitman, Phys. Rev. D \textbf{53},
7162 (1996).

\bibitem{Nik70} A.I. Nikishov, Sov. Phys. JETP \textbf{30}, 660 (1970); S.P.
Gavrilov, D.M. Gitman and A.E. Gonçalves, J. Math. Phys. \textbf{39}, 3547
(1998).

\bibitem{Mad08} J. Madsen, Phys. Rev. Lett. \textbf{100}, 151102 (2008).

\bibitem{CasNN79} A. Casher, H. Neuberger, and S. Nussinov, Phys. Rev. D 
\textbf{20,} 179 (1979); E.G. Gurvich, Phys. Lett. B \textbf{87,} 386 (1979).

\bibitem{KhaLT06} D. Kharzeev, E. Levin, and K. Tuchin, Phys. Rev. C \textbf{%
75,} 044903 (2007); T. Lappi and L. McLerran, Nucl. Phys. A \textbf{772, }%
200 (2006).

\bibitem{GavGT06} S.P. Gavrilov, D.M. Gitman, and J.L. Tomazelli, Nucl.
Phys. B \textbf{795, }645 (2008).

\bibitem{NayN05} G.C. Nayak, and P. van Nieuwenhuizen, Phys. Rev. D \textbf{%
71,} 125001 (2005); G.C. Nayak, Phys. Rev. D \textbf{72,} 125010 (2005); F.
Cooper and G.C. Nayak, Phys. Rev. D \textbf{73,} 065005 (2006).
\end{thebibliography}
\end{document}